\documentclass[10pt,english,aps,prd,showkeys,showpacs,notitlepage]{revtex4-1}
\usepackage[T1]{fontenc}
\usepackage[utf8]{inputenc}
\usepackage{array}
\usepackage{float}
\usepackage{booktabs}
\usepackage{amsmath}
\usepackage{amssymb}
\usepackage{silence}
\makeatletter

\providecommand{\tabularnewline}{\\}

 \@ifundefined{textcolor}{}
 {%
   \definecolor{BLACK}{gray}{0}
   \definecolor{WHITE}{gray}{1}
   \definecolor{RED}{rgb}{1,0,0}
   \definecolor{GREEN}{rgb}{0,1,0}
   \definecolor{BLUE}{rgb}{0,0,1}
   \definecolor{CYAN}{cmyk}{1,0,0,0}
   \definecolor{MAGENTA}{cmyk}{0,1,0,0}
   \definecolor{YELLOW}{cmyk}{0,0,1,0}
 }

\usepackage{graphicx}
\usepackage{tikz}

\makeatother

\usepackage{babel}
\begin{document}

\title{Quark horizontal flavor symmetry and two-Higgs doublet in (7+1)-dimensional
extended spin space}

\author{R. Romero}
\email{romero@ciencias.unam.mx}

\altaffiliation{currently at Departamento de Física, Universidad Autónoma Metropolitana-Iztapalapa, Ciudad de México 09340, México.}

\author{J. Besprosvany}
\email{bespro@fisica.unam.mx}

\selectlanguage{english}%

\affiliation{Instituto de Física, Universidad Nacional Autónoma de México, Apartado
Postal 20-364, Ciudad de México 01000, México}
\begin{abstract}
An extended spin-space model in $7+1$ dimensions is presented that
describes the standard-model electroweak quark sector. Up to four generations
of massless and massive quarks and two-Higgs doublets derive from  the associated representation space, in addition to the W- and Z-vector bosons.
 Other mass operators are obtained that put restrictions on additional non-Higgs scalars  and their vacuum 
 expectation value.   
After symmetry breaking, the scalar  components give rise to a  hierarchy effect vertically (within doublets) associated to the  Higgs fields, and horizontally (within generations)
 associated to the non-Higgs elements.
\end{abstract}
\maketitle

\section{introduction}

The Standard Model (SM) of particle physics is very successful in
describing the fundamental particles and their interactions, but it
is also phenomenological, requiring a large amount of experimental
input such as fermion masses, coupling constants and mixing angles.
The recent discovery of a Higgs-like particle at the LHC further validates
the SM, but the absence of positive results in new particles searches
\citep{Craig:2013cxa,ANDP:ANDP201500162} also severely constrains
favored beyond the SM proposals, such as supersymmetry (SUSY)\citep{Aaboud2016,PhysRevD.94.032003,Khachatryan2016152,Khachatryan2016}.

Two elements frequently used in beyond-the-SM model building,  
that are within reach of experimental verification at the Large Hadron Collider (LHC) in
upcoming years, are the extension of the Higgs sector, as in two-Higgs
doublet (2HD) models\citep{Branco:2011iw}, and the inclusion of horizontal
(flavor) symmetries \citep{Ponce:1995wx,Fritzsch:1999ee,rosner2001flavor,Babu:2009fd}.
The latter can be global or gauged, and the flavor group can be either
discrete or a continuous Lie group. In trying to explain the origin
of the mass hierarchies and mixings of quarks and leptons, it has
been proposed that the flavor symmetry might be broken by Froggatt-Nielsen
scalars, or flavons\citep{FROGGATT1979277}, which are scalars transforming
non-trivially under the flavor group that couple to fermions and acquire
vacuum expectation values. Different choices for the flavor group,
as well as the breaking scale and mechanism, lead to different phenomenological
outcomes.

One common feature of models incorporating the above elements is that the
additional Higgs structure and/or the flavor symmetry are proposed
\emph{ab initio}, rather than derived. To mention some recent examples,
in Ref. \onlinecite{Bauer2015} a 2HD model is proposed where
the two Higgses serve as the flavon and the flavor breaking scale
is just the electroweak one. In Ref. \onlinecite{PhysRevD.91.075006}
a 2HD and additional scalar singlets are used in conjunction with
a gauged $\text{U}(1)$ horizontal symmetry to explain the SM deviations
in  B-meson  decays found by the LHCb run, while in Ref. \onlinecite{PhysRevD.92.095013}
an extension of a 2HD model is considered in the context of the Dine-Fischler-Srednicki-Zhitnitsky
(DFSZ) axion model\citep{DINE1981199,Zhitnitsky:1980tq}.

In looking for model-building alternatives, and motivated by the original
Kaluza-Klein idea of  gauge-interactions and gravity unification through the addition of more spacetime
dimensions, it has been proposed\citep{Besprosvany:2010zz,Besprosvany2015199}
to enlarge instead the dimensions of the abstract vector space in
which spin-$1/2$ objects reside, commonly referred to as spin space,
but keeping Lorentz symmetry in the standard four-dimensional spacetime.
Hence, the extra dimensions are only relevant in spin space, while
they are \emph{frozen} in spacetime. The motivation for doing so is
that, regarding the classification of physical fields by their transformation
properties under the Poincaré group, the spin-$1/2$ representation
is the fundamental one, since all tensor representations can be obtained
from the former, but not the other way around. In another interpretation,
one proceeds by checking additional discrete spaces, described in
terms of a Clifford algebra and independently of the configuration-space description, that allow for the inclusion of the relevant states
and operators\citep{Besprosvany:2010zz}. In this respect, a 
(7+1)-dimensional[d]  discrete space is the minimum space that allows for the
description of different flavor quarks and the operators that classify
them, since lower dimensional spaces have been studied and found not
big enough for that purpose\citep{Besprosvany:2002zr,Besprosvany2001323,Besprosvany:2002py,Besprosvany:2010zz}.

Properties and results of the extended spin model, within the context
of relativistic quantum mechanics, have been studied before in various
dimensions\citep{Besprosvany:2002zr,Besprosvany:2002py,Besprosvany:2002tv,Besprosvany:2003jx},
and recently, a formal translation was presented between fields and
Lagrangians in the extended spin space and the conventional formulation,
and in particular for the SM. Thus, a direct interpretation of the
model in terms of a field theory is now available. Given that the
extended space is constructed in terms of ket-bras constructed from elementary doublets,
a matrix-vector space results, which is suitably endowed with a metric,
and is describable in terms of a Clifford algebra. Also, maintaining
the 4-d Lorentz symmetry results in additional operators in the same
matrix space that are associated to gauge and flavor symmetries; as
these commute with the former, the Coleman-Mandula theorem\citep{Coleman:1967ad}
is satisfied. The latter makes it possible for symmetry operators,
both gauge and Lorentz, to be elements of the same matrix space, along
with the physical fields they classify. 

In this paper, we derive a model in the $\left(7+1\right)$-d extended
spin space, where we obtain up to four massless and massive quark
generations, two Higgs doublets, and both horizontal and vertical
mass relations. There are also  mass operators   that put restrictions on additional
non-Higgs scalars  and their vacuum 
 expectation value.  We show they constrain   the quark mass spectrum; in particular,  they
   generate linearly vertical and horizontal hierarchies. Thus, they play a  similar role  to flavons, with differences. 
 All these features are obtained from the model under standard
physical assumptions. The dimension of the space is chosen because it
is the minimal one required for massive quarks, since the previous
lower $\left(5+1\right)$-d space was studied\citep{Besprosvany2001323,Besprosvany:2002py},
and found not big enough to accommodate left- and right-handed quark
fields with the SM quantum numbers. The paper is organized as follows:
In section II, general properties of the model are reviewed, including
operators, symmetry transformations, and field and Lagrangian representation. Section III presents
the $\left(7+1\right)$-d model, giving the classification of operators
and states, and obtaining massless and massive quarks, and  Higgs-like particles, and non-Higgs scalars, along with hierarchies,
both horizontal and vertical.  In   Section IV,
we give concluding remarks.

\section{Extended spin model}

Let us begin by providing a brief introduction to the features of the extended spin model.

\subsection{General Properties}

Let us consider an $N$-dimensional Clifford algebra $\mathcal{C}_{N}$,
with $N$ even, generated by a set of $N$ gamma matrices of dimension
$2^{N/2}\times2^{N/2}$ obeying the defining property of the algebra\citep{Snygg1997,Benn1987}

\begin{equation}
\gamma_{\alpha}\gamma_{\beta}+\gamma_{\beta}\gamma_{\alpha}=2g_{\alpha\beta},\label{eq:1}
\end{equation}

\noindent where $g_{\alpha\beta}$ is the metric tensor with signature $(+,-,...,-)$
and $\alpha,\beta=0,1,\ldots3,5,\ldots,N$\footnote{Following standard practice, the label 4 is omitted.}.
The gamma matrices are taken with the standard Hermiticity properties:

\begin{equation}
\begin{array}{rc}
\gamma_{0}^{\dagger}=\gamma_{0},\\
\gamma_{i}^{\dagger}=-\gamma_{i} & i=1,\ldots,N.
\end{array}\label{eq:1-1}
\end{equation}

\noindent The $N$ gamma matrices and all their linearly independent
products form a set of $2^{N}$ elements, which constitutes a basis
for the vector space of complex $2^{N/2}\times2^{N/2}$ matrices.
The spin Lorentz generators and finite Lorentz transformations acting
on spinors have standard expressions in the four dimensional Clifford
algebra $\mathcal{C}_{4}$, namely

\begin{equation}
\begin{array}{ccc}
\sigma_{\mu\nu}=\frac{i}{2}\left[\gamma_{\mu},\gamma_{\nu}\right] & \textrm{with} & \mu,\nu=0,\ldots,3\end{array},\label{eq:2}
\end{equation}

\noindent 
\begin{equation}
S(\Lambda)=e^{-\frac{i}{4}\sigma_{\mu\nu}\omega^{\mu\nu}}.\label{eq:3}
\end{equation}

\noindent As a result, the $(3+1)$-d  gamma matrices transform under $S(\Lambda)$
as vectors

\begin{equation}
S(\Lambda)\gamma^{\mu}S(\Lambda)^{-1}=\left(\Lambda^{-1}\right)^{\mu}\,_{\nu}\gamma^{\nu},\label{eq:10-1}
\end{equation}

\noindent $\mu,\nu=0,\ldots,3$, while the remaining $N-4$ gamma
matrices $\gamma_{a}$, $a=4,\ldots,N-1$, and their products commute
with $\sigma_{\mu\nu}$, e.g. $\left[\gamma_{6},\sigma_{01}\right]=i\gamma_{6}\gamma_{0}\gamma_{1}-i\gamma_{0}\gamma_{1}\gamma_{6}=0$,
so they are indeed Lorentz scalars

\begin{equation}
S(\Lambda)\gamma^{a}S(\Lambda)^{-1}=\gamma^{a}.\label{eq:10-2}
\end{equation}

\noindent These scalars can be identified with generators of continuous
symmetries, either gauge or global, and they automatically satisfy
the Coleman-Mandula theorem\citep{Coleman:1967ad} since they commute
with the Lorentz generators. This latter feature allows for  the inclusion
of spacetime and gauge groups generators in the same space, and provides
an alternative to a graded Lie algebra for evading the Coleman-Mandula
theorem, as is done in SUSY. Furthermore, the states on which the
operators act are also elements of the space, so we have a matrix
space with the dimensionality restricting the type and number of allowed
scalar symmetries. The scalars generate the unitary-group combination

\begin{equation}
\mathcal{S}_{N-4}=\frac{1}{2}(I+\tilde{\gamma}_{5})\textrm{U}\left(2^{(N-4)/2}\right)\oplus\frac{1}{2}(I-\tilde{\gamma}_{5})\textrm{U}\left(2^{(N-4)/2}\right),\label{eq:4}
\end{equation}

\noindent where $I$ stands for the $N$-d identity matrix and $\tilde{\gamma}_{5}$
is the 4-d chirality matrix, defined as

\noindent 
\begin{equation}
\tilde{\gamma}_{5}\equiv-i\gamma_{0}\gamma_{1}\gamma_{2}\gamma_{3}.\label{eq:4-1}
\end{equation}

\noindent Given that $\textrm{U}\left(2^{(N-4)/2}\right)$ possesses
$2^{N-4}$ generators, the number of elements in $\mathcal{S}_{N-4}$
is $2^{N-3}$.

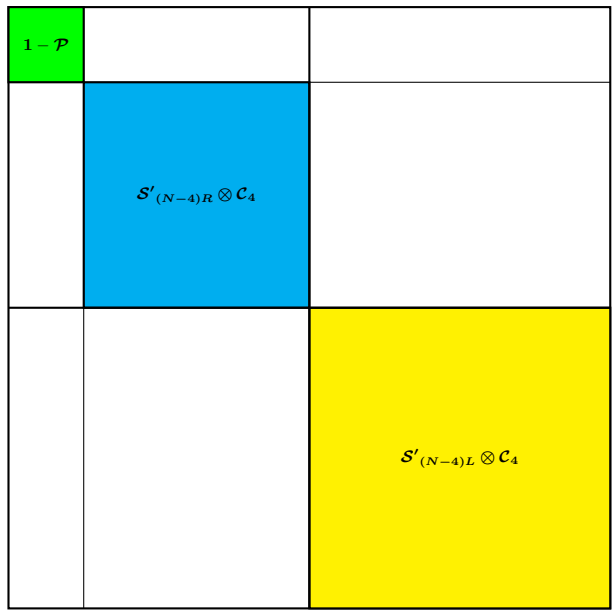
\begin{figure}[t]
\begin{centering}
\begin{tikzpicture}[scale=1,every node/.style={scale=0.7}] 
\draw[thick] (0,0) rectangle (8,8);

\draw [fill=yellow,thick] (4,0) rectangle (8,4); 
\draw [fill=cyan,thick] (1,4) rectangle (4,7); 
\draw [fill=green,thick] (0,7) rectangle (1,8); 

\draw[thick] (4,0) -- (4,8); 
\draw[thick] (0,4) -- (8,4);
\draw (1,0) -- (1,8);
\draw (0,7) -- (8,7);
\node at (0.5,7.5) {$\boldsymbol{1}-\boldsymbol{\mathcal{P}}$};
\node at
(2.5,5.5) {$\boldsymbol{\mathcal{S'}}_{\boldsymbol{(N-4)R}}\boldsymbol{\otimes}\boldsymbol{\mathcal{C}}_{\boldsymbol{4}}$};
\node at
(6,2) {$\boldsymbol{\mathcal{S'}}_{\boldsymbol{(N-4)L}}\boldsymbol{\otimes}\boldsymbol{\mathcal{C}}_{\boldsymbol{4}}$};

\end{tikzpicture}
\par\end{centering}
\caption{Schematic representation of symmetry generators in extended spin space, arranged in chiral blocks as in Eq. (\ref{eq:4}), producing both scalar and Lorentz generators\citep{Besprosvany:2014lwa}, and $\mathcal{S}'$ defined in Eq. \ref{eq:11}. The operator $\mathcal{P}$ defines the way in which the space is divided. }

\end{figure}

\begin{figure}[t]
\begin{centering}
\begin{tikzpicture}[scale=1,every node/.style={scale=0.7}] 
\draw[thick] (0,0) rectangle (8,8);

\draw [fill=yellow,thick] (1,4) rectangle (4,7); 
\draw [fill=yellow,thick] (4,0) rectangle (8,4); 
\draw [fill=cyan,thick] (1,0) rectangle (4,4); 
\draw [fill=cyan,thick] (4,4) rectangle (8,7); 
\draw [fill=lime,thick] (0,0) rectangle (1,4); 
\draw [fill=lime,thick] (0,4) rectangle (1,7);
\draw [fill=lime,thick] (1,7) rectangle (4,8);
\draw [fill=lime,thick] (4,7) rectangle (8,8);
\draw [fill=orange,thick] (0,7) rectangle (1,8); 

\draw[thick] (4,0) -- (4,8); 
\draw[thick] (0,4) -- (8,4);
\draw (1,0) -- (1,8);
\draw (0,7) -- (8,7);
\node at (0.5,7.5) {$\boldsymbol{1}-\boldsymbol{\mathcal{P}}$};
\node at (0.5,5.5) {\textbf{F}};
\node at (0.5,2) {\textbf{F}};
\node at (2.5,5.5) {\textbf{V}};
\node at (2.5,7.5) {$\overline{\mathbf{F}}$};
\node at (2.5,2) {\textbf{S},\textbf{A}};
\node at (6,2) {\textbf{V}};
\node at (6,5.5) {\textbf{S},\textbf{A}};
\node at (6,7.5) {$\overline{\mathbf{F}}$};

\end{tikzpicture}
\par\end{centering}
\caption{Representation of states in extended spin space\citep{Besprosvany:2014lwa},
classified according to their Lorentz transformation properties: fermion
(F), vector (V), scalar (S), and anti-symmetric tensor (A). Antifermions
($\bar{\text{F}}$) correspond to the Hermitian conjugate, and the
blocks for V, S and A also contain antiparticle solutions.}
\end{figure}

\subsection{Operators and symmetry transformations }

Physical fields are associated with elements of $\mathcal{C}_{N}$,
classified by operators from $\mathcal{C}_{4}\otimes\mathcal{S}_{N-4}$,
so in general, a field $\psi$ has the structure

\begin{equation}
\left(\mbox{elements of 3+1 space }\right)\times\left(\begin{array}{c}
\text{combination of products }\\
\text{of elements of \ensuremath{\mathcal{S}_{N-4}}}
\end{array}\right).\label{eq:10}
\end{equation}

\noindent Symmetry transformations act on fields as

\begin{equation}
\Psi(x)\rightarrow U\Psi(x)U^{\dagger},\label{eq:5}
\end{equation}

\noindent and commutators are used for the evaluation of operators

\begin{equation}
\left[\mathcal{O},\Psi(x)\right]=\lambda\Psi(x).\label{eq:6}
\end{equation}

\noindent This last equation is to be interpreted as an eigenvalue
equation, that is, $\psi$ is an eigenstate of $\mathcal{O}$ with
eigenvalue $\lambda$. The inner product of two fields is defined
accordingly to a matrix space

\begin{equation}
\left\langle \phi\mid\Psi\right\rangle =\text{tr}\left(\phi{}^{\dagger}\Psi\right).\label{eq:6-1}
\end{equation}

\noindent Vector and scalar fields transform respectively under Eq.
(\ref{eq:3}) as the matrices $\gamma^{\mu}$, $\mu=0,\ldots,3$ and
$\gamma^{a}$, $a=5,\ldots,N$, Eqs. (\ref{eq:10-1}) and (\ref{eq:10-2}).

 For the fermion representation, a projector operator $\mathcal{P}$,
obtained from elements of $\mathcal{S}$, is used such that it acts
on both the Poincaré generators $\mathcal{J}_{\mu\nu}=i\left(x_{\mu}\partial_{\nu}-x_{\nu}\partial_{\mu}\right)+\frac{1}{2}\sigma_{\mu\nu}$
and the $\mathcal{S}_{N-4}$ symmetry operator space

\begin{equation}
\begin{array}{c}
\mathcal{J}'_{\mu\nu}=\mathcal{P}\mathcal{J}_{\mu\nu}=\mathcal{P}\left[i\left(x_{\mu}\partial_{\nu}-x_{\nu}\partial_{\mu}\right)+\frac{1}{2}\sigma_{\mu\nu}\right],\\
\\
\mathcal{S}'=\mathcal{P}\mathcal{S}.
\end{array}\label{eq:11}
\end{equation}
A fermion field is then required to include $1-\mathcal{P}$
in the general structure of Eq. (\ref{eq:10}), so that Poincaré and
gauge generators act trivially on its rhs when evaluating commutators
as in Eq. (\ref{eq:6}), since $(1-\mathcal{P})\mathcal{P}=0$. This
leads to fermions transforming in the fundamental representation of
both the Poincaré and gauge groups, that is, for those groups instead
of Eq. (\ref{eq:5}) we have

\begin{equation}
\Psi\rightarrow U\Psi,\label{eq:12-1}
\end{equation}

\noindent where $\Psi$ stands for a fermion field. Flavor operators,
on the contrary, act non-trivially from the right-hand side and give zero from the left, as will
be shown in the next section. This is heuristically understood by thinking of $\Psi$ as the matrix $\Psi\sim\left|\psi\right\rangle\left\langle a\right|$, with the ket $\left|\psi\right\rangle$ carrying Lorentz and gauge group information, while the bra $\left\langle a\right|$ carries information about the flavor.

 Fig. 1   presents   schematically   symmetry operators in the matrix space, and Fig. 2 shows the corresponding  Lorentz states: scalars, vectors, fermions, and anti-symmetric tensors, arranged in the same matrix space.

\subsection{Lagrangian formulation}

Interactive Lagrangians\citep{Besprosvany:2014lwa} can be given in
terms of vector, scalar and fermion fields conforming to the general
structure of Eq. (\ref{eq:11}). These fields are:

\subsection*{Vector field}

\begin{equation}
A_{\mu}^{a}(x)\gamma_{0}\gamma_{\mu}I_{a},\label{eq:14}
\end{equation}

\noindent where $\gamma_{0}\gamma_{\mu}\in\mathcal{C}_{4}$ and $I_{a}\in\mathcal{S}'_{n-4}$
is a generator of a given unitary group. 

\subsection*{Scalar field}

\begin{equation}
\phi^{a}(x)\gamma_{0}\Gamma_{a}^{S},\label{eq:15-1}
\end{equation}

\noindent with $\Gamma_{a}^{S}$ an element of $\mathcal{S}$.

\subsection*{Fermion field}

\begin{equation}
\psi_{\alpha}^{a}(x)L^{\alpha}P_{F}\Gamma_{a}^{F},\label{eq:16-1}
\end{equation}

\noindent where $\Gamma_{a}^{F}$ is an element of $\mathcal{S}$,
and $L^{\alpha}$ represents a spin polarization component, e.g. $L^{1}=\left(\gamma_{1}+i\gamma_{2}\right)$.
The operator $P_{F}$ is a projection operator of the type in Eq.
(\ref{eq:11}) such that

\begin{equation}
P_{F}\gamma^{\mu}=\gamma^{\mu}P_{F}^{c},\label{eq:17-1}
\end{equation}

\noindent with $P_{F}^{c}=1-P_{F}$. Vector and scalar fields transform
under gauge transformations as in Eq. (\ref{eq:5}), while fermions
transform as in Eq. (\ref{eq:12-1}), with $U$ given by

\begin{equation}
U=\exp\left[-iI_{a}\alpha_{a}(x)\right].\label{eq:17-2}
\end{equation}

\noindent For example, a gauge-invariant fermion-vector Lagrangian
is given by

\begin{equation}
\dfrac{1}{N_{f}}\text{tr}\Psi^{\dagger}\left\{ \left[i\partial_{\mu}-gA_{\mu}^{a}(x)I_{a}\right]\gamma^{0}\gamma^{\mu}-m\gamma^{0}\right\} \Psi,\label{eq:18-1}
\end{equation}

\noindent where $\Psi$ is a fermion field as in Eq. (\ref{eq:16-1}), $g$ is the coupling constant, and $N_{f}$ contains the normalization. 

\section{7+1-dimensional model}

We now present a $7+1$ dimensional model of the electroweak quark sector in extended space.

\subsection{Generators and operators}

The $\left(7+1\right)$-d Clifford algebra is generated by the eight
$16\times16$ matrices

\begin{equation}
\gamma_{0},\gamma_{1},\ldots,\gamma_{8}.\label{eq:12}
\end{equation}

\noindent The first four matrices form the Lorentz generators $\sigma_{\mu\nu}$
of Eq. (\ref{eq:2}), and the set of unitary scalars consists of 32 elements
that generate the group combination ${\textstyle \mathcal{S}=P_{+}\textrm{U}\left(4\right)\oplus P_{-}\textrm{U}\left(4\right)}$,
with $P_{\pm}=\frac{1}{2}\left(1\pm\hat{\gamma}_{5}\right)$ and $\tilde{\gamma}_{5}$
the chirality matrix within 4-d, defined in Eq. (\ref{eq:4-1}).
The 32 elements of $\mathcal{S}$ are: four matrices $\gamma_{a}$,
$a=5,\ldots,8$, six pairs $\gamma_{ab}\equiv\gamma_{a}\gamma_{b}$,
$a<b$, four triplets $\gamma_{abc}\equiv\gamma_{a}\gamma_{b}\gamma_{c}$,
and one quadruplet $\gamma_{5}\gamma_{6}\gamma_{7}\gamma_{8}$. To
these fifteen matrices we add the same matrices multiplied by $\tilde{\gamma}_{5}$,
plus $\tilde{\gamma}_{5}$ itself and the identity to complete the
32 scalars.

The maximal number of elements in $\mathcal{S}$ that commute with
each other is eight, and these elements define the eight-dimensional
Cartan subalgebra $\mathfrak{h}$ of $\mathcal{S}$, for which we
choose the basis

\begin{equation}
\begin{array}{cccccccc}
1, & \hat{\gamma}_{5}, & \gamma_{5}\gamma_{6}, & \gamma_{7}\gamma_{8}, & \gamma_{5}\gamma_{6}\gamma_{7}\gamma_{8}, & \gamma_{5}\gamma_{6}\tilde{\gamma}_{5}, & \gamma_{7}\gamma_{8}\tilde{\gamma}_{5}, & \gamma_{5}\gamma_{6}\gamma_{7}\gamma_{8}\tilde{\gamma}_{5}.\end{array}\label{eq:13}
\end{equation}

\noindent This basis is used to build operators that classify the
particles according to their quantum numbers, seeking a configuration
as close as possible to the SM. One degree of freedom in $\mathfrak{h}$
is assigned to the projector $\mathcal{P}$, which characterizes
fermions, as in Eq. (\ref{eq:11}). Two other are the  $SU(2)_{L}\times U(1)_{Y}$ 
gauge-group generators that can be chosen diagonal,
 namely,
the third component of isospin $I_{3}$ and the hypercharge $Y$.
Of the remaining five, three are associated to flavor, to be described
below, and the rest are just the identity and the chirality.

To obtain the diagonal operators let us recast the basis in Eq. (\ref{eq:13})
in terms of the projection operators

\begin{equation}
\begin{array}{cc}
P_{R1}= & \frac{1}{8}(1+\tilde{\gamma}_{5})(1+i\gamma_{5}\gamma_{6})(1+i\gamma_{7}\gamma_{8}),\\
P_{R2}= & \frac{1}{8}(1+\tilde{\gamma}_{5})(1+i\gamma_{5}\gamma_{6})(1-i\gamma_{7}\gamma_{8}),\\
P_{R3}= & \frac{1}{8}(1+\tilde{\gamma}_{5})(1-i\gamma_{5}\gamma_{6})(1+i\gamma_{7}\gamma_{8}),\\
P_{R4}= & \frac{1}{8}(1+\tilde{\gamma}_{5})(1-i\gamma_{5}\gamma_{6})(1-i\gamma_{7}\gamma_{8}),\\
P_{L1}= & \frac{1}{8}(1-\tilde{\gamma}_{5})(1+i\gamma_{5}\gamma_{6})(1+i\gamma_{7}\gamma_{8}),\\
P_{L2}= & \frac{1}{8}(1-\tilde{\gamma}_{5})(1+i\gamma_{5}\gamma_{6})(1-i\gamma_{7}\gamma_{8}),\\
P_{L3}= & \frac{1}{8}(1-\tilde{\gamma}_{5})(1-i\gamma_{5}\gamma_{6})(1+i\gamma_{7}\gamma_{8}),\\
P_{L4}= & \frac{1}{8}(1-\tilde{\gamma}_{5})(1-i\gamma_{5}\gamma_{6})(1-i\gamma_{7}\gamma_{8}),
\end{array}\label{eq:14-1}
\end{equation}

\noindent which are schematically shown in the matrix space in Fig. 3. Each one of these projectors has two degenerate non-zero
eigenvalues, set to one, and all scalar diagonal operators can be written
as their linear combination. For the  right-handed
quarks description, we need two different hypercharge quantum numbers, namely $4/3$
and $-2/3$,  respectively, for    up- and down-type quarks, 
so the right-handed part of the hypercharge operator must be a linear
combination of at least two of the first four operators in Eq. (\ref{eq:14-1}),
with the weights given by the two eigenvalues. If the operators
$P_{R1}$ and $P_{R2}$   make room for flavor and
$1-\mathcal{P}$, necessary for the description of different fermion families,
this leaves $P_{R3}$, and $P_{R4}$ as the only candidates,
and we assign

\begin{equation}
\frac{1}{2}(1+\tilde{\gamma}_{5})\hat{Y}=\frac{4}{3}P_{R3}-\frac{2}{3}P_{R4},\label{eq:15-3}
\end{equation}

\noindent for the right-handed part of the hypercharge operator. (Alternatively, fixing $P_{R3}$ and $P_{R4}$ with Eq. (\ref{eq:15-3}), leaves  $P_{R1}$ and $P_{R1}$ for $1-\mathcal{P}$ and
flavor association.) The
choice of the two operators in Eq. (\ref{eq:15-3}) is not unique,
with different matrix operator representations of  and states, leading
to the same physics.

\noindent 
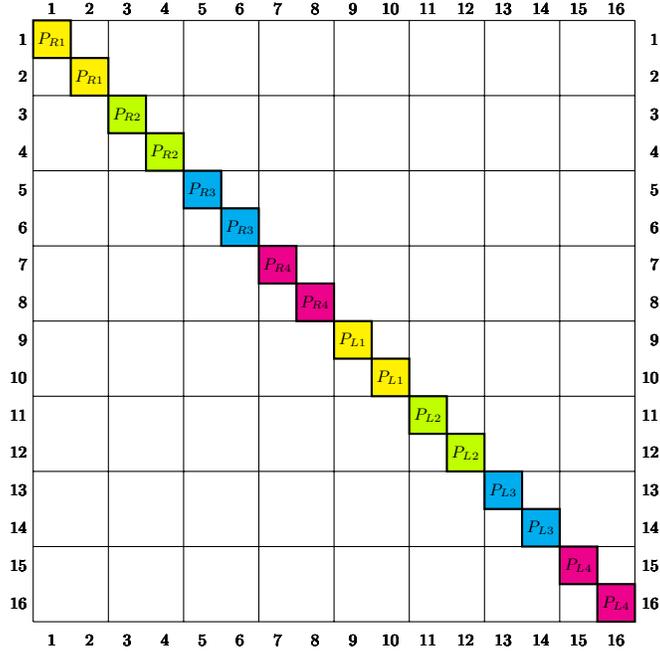
\begin{figure}[t]
\begin{centering}
\begin{tikzpicture}[scale=1,every node/.style={scale=0.7}] 
\draw (0,0) grid (8,8); 

\draw [fill=yellow,thick] (0,7.5) rectangle (0.5,8);
\draw [fill=yellow,thick] (0.5,7) rectangle (1,7.5);
\draw [fill=lime,thick] (1,6.5) rectangle (1.5,7);
\draw [fill=lime,thick] (1.5,6) rectangle (2,6.5);
\draw [fill=cyan,thick] (2,5.5) rectangle (2.5,6);
\draw [fill=cyan,thick] (2.5,5) rectangle (3,5.5);
\draw [fill=magenta,thick] (3,4.5) rectangle (3.5,5);
\draw [fill=magenta,thick] (3.5,4) rectangle (4,4.5);

\draw [fill=yellow,thick] (4,3.5) rectangle (4.5,4);
\draw [fill=yellow,thick] (4.5,3) rectangle (5,3.5);
\draw [fill=lime,thick] (5,2.5) rectangle (5.5,3);
\draw [fill=lime,thick] (5.5,2) rectangle (6,2.5);
\draw [fill=cyan,thick] (6,1.5) rectangle (6.5,2);
\draw [fill=cyan,thick] (6.5,1) rectangle (7,1.5);
\draw [fill=magenta,thick] (7,0.5) rectangle (7.5,1);
\draw [fill=magenta,thick] (7.5,0) rectangle (8,0.5);


\draw (1,6.5) -- (2,6.5);
\draw (1.5,6) -- (1.5,7);
\draw (3,4.5) -- (4,4.5);
\draw (3.5,4) -- (3.5,5);
\draw (5,2.5) -- (6,2.5);
\draw (5.5,2) -- (5.5,3);
\draw (7,0.5) -- (8,0.5);
\draw (7.5,0) -- (7.5,1);
\draw (0,7.5) -- (1,7.5);
\draw (0.5,7) -- (0.5,8);
\draw (2,5.5) -- (3,5.5);
\draw (2.5,5) -- (2.5,6);
\draw (4,3.5) -- (5,3.5);
\draw (4.5,3) -- (4.5,4);
\draw (6,1.5) -- (7,1.5);
\draw (6.5,1) -- (6.5,2);


\node at (1.25,6.75) {$P_{R2}$}; 
\node at (1.75,6.25) {$P_{R2}$}; 
\node at (3.25,4.75) {$P_{R4}$};
\node at (3.75,4.25) {$P_{R4}$};
\node at (5.25,2.75) {$P_{L2}$};
\node at (5.75,2.25) {$P_{L2}$};
\node at (7.25,0.75) {$P_{L4}$};
\node at (7.75,0.25) {$P_{L4}$};
\node at (0.25,7.75) {$P_{R1}$};
\node at (0.75,7.25) {$P_{R1}$};
\node at (2.25,5.75) {$P_{R3}$};
\node at (2.75,5.25) {$P_{R3}$};
\node at (4.25,3.75) {$P_{L1}$};
\node at (4.75,3.25) {$P_{L1}$};
\node at (6.25,1.75) {$P_{L3}$};
\node at (6.75,1.25) {$P_{L3}$};


\foreach \i/\valorI in {1/1,2/3,3/5,4/7,5/9,6/11,7/13,8/15}
\foreach \j/\valorP in {1/2,2/4,3/6,4/8,5/10,6/12,7/14,8/16}  
{     
\node[anchor=south] at (\i-0.75,8) {$\valorI$}; 
\node[anchor=south] at (\j-0.25,8) {$\valorP$};    
\node[anchor=east] at (0,-\i+8.75) {$\valorI$};
\node[anchor=east] at (0,-\j+8.25) {$\valorP$};
\node[anchor=south] at (\i-0.75,-0.4) {$\valorI$}; 
\node[anchor=south] at (\j-0.25,-0.4) {$\valorP$};   
\node[anchor=east] at (8.4,-\i+8.75) {$\valorI$};
\node[anchor=east] at (8.4,-\j+8.25) {$\valorP$};
} 
\end{tikzpicture}
\par\end{centering}
\caption{Matrix representation of the Cartan basis (cf. eq. (\ref{eq:14-1}))
in extended spin space in 7+1 dimensions. The eight-dimensional basis
is represented here  in terms of the projection operators $P_{R,Li}$,
$i=1,\ldots,4.$ The subscripts $R,L$ refer
to the chirality: $R$ for operators containing $1+\tilde{\gamma}_{5}$
(right-handed), and $L$ for operators containing $1-\tilde{\gamma}_{5}$
(left-handed).}
\end{figure}

The left-handed part of the hypercharge operator is symmetrically
given as

\begin{equation}
\frac{1}{2}(1-\tilde{\gamma}_{5})Y=\frac{1}{3}\left(P_{L3}+P_{L4}\right),\label{eq:30}
\end{equation}

\noindent where both weights equal $1/3$ as such is the hypercharge
eigenvalue for both up- and down-type left-handed quarks. Thus, the
complete operator is 

\begin{gather}
\begin{split}Y & =\frac{1}{3}\left(4P_{R3}-2P_{R4}+P_{L3}+P_{L4}\right),\\
 & =\frac{1}{6}\left(1-i\gamma_{5}\gamma_{6}\right)\left(1+i\frac{3}{2}(1+\tilde{\gamma}_{5})\gamma_{7}\gamma_{8}\right).
\end{split}
\label{eq:31}
\end{gather}

Tracing $Y$,  operator $\mathcal{P}$  in Eq. (\ref{eq:11}) must be given by
a linear combination of the same operators appearing in the hypercharge
expression, Eq. (\ref{eq:31}), multiplied by the same, positive weight
$\omega$. This leads to 

\begin{equation}
\mathcal{P}=\omega(P_{R3}+P_{R4}+P_{L3}+P_{L4}),\label{eq:32}
\end{equation}

\noindent 
where $\omega=1$ to have $\mathcal{P}\mathcal{P}=\mathcal{P}$.
 $\mathcal{P}$  provides fermions with standard transformation
properties, as stated in Eqs. (\ref{eq:11}) and (\ref{eq:12-1}).
As a consequence, if $\Phi$ represents a scalar, vector, or in general,
a tensor field, we have

\begin{equation}
\left[\mathcal{P},\Phi\right]=0,\label{eq:16-2}
\end{equation}

\noindent while for a fermion field $\Psi$, either massless or massive,
we get

\begin{equation}
\left[\mathcal{P},\Psi\right]=\lambda\Psi,\label{eq:17-3}
\end{equation}

\noindent and $-\lambda$ for $\Psi^{\dagger}$. Hence, $\mathcal{P}$
is the generator of a $\text{U}(1)$ symmetry for fermions, which
we associate with baryon number, so from now on $\mathcal{P}$ will
be referred to as the baryon number operator $B$, with $\lambda_{B}=1/3$.
In terms of the gamma matrices, the operator is

\begin{equation}
B=\frac{1}{6}(1-i\gamma_{5}\gamma_{6})=\dfrac{1}{3}(P_{R3}+P_{R4}+P_{L3}+P_{L4}).\label{eq:29}
\end{equation}

\noindent It is also possible to start by initially determining the
$B$ operator, requiring the spectrum to contain fermions with standard
transformation properties under the Lorentz and gauge groups. This
approach implies that $B$ must be given by a linear
combination of four operators from Eq. (\ref{eq:14-1}), two right-handed
and two left-handed. Again, different choices change the matrix structure
but not the physics. Also, this   baryon-number definition  is not
related to the remnant $\textrm{U}(1)$ symmetry of QCD after chiral
symmetry breaking, and in this respect it is not dynamical, but rather,
it follows from the choice of the projector $\mathcal{P}$. In other words,
once the projector is chosen to allow for the existence of fermions
with standard transformation properties, the baryon-number definition
follows naturally.

Given that the weak isospin generators commute with $Y$, we
can construct the diagonal generator $I_{3}$ with the same left-handed
projectors used in $Y$, Eq. (\ref{eq:30}), and we assign

\begin{equation}
I_{3}=\frac{1}{2}(P_{L3}-P_{L4})=\frac{i}{8}(1-\tilde{\gamma}_{5})(1-i\gamma_{5}\gamma_{6})\gamma_{7}\gamma_{8}.\label{eq:33}
\end{equation}

\noindent The rest of the $\textrm{SU}(2)_{L}$ generators are obtained
from the non-diagonal elements of $\mathcal{S}$, by requiring that
they fulfill the group Lie algebra $[I_{k},I_{l}]=i\varepsilon_{klm}I_{m}$,
and that they also commute with $Y$. They are given by 

\begin{equation}
\begin{split}I_{1} & ={\displaystyle \frac{i}{8}(1-\tilde{\gamma}_{5})(1-i\gamma_{5}\gamma_{6})\gamma^{7}},\\
I_{2} & {\displaystyle =\frac{i}{8}(1-\tilde{\gamma}_{5})(1-i\gamma_{5}\gamma_{6})\gamma^{8}},
\end{split}
\label{eq:16}
\end{equation}

\noindent Finally, the charge operator $Q$ is given by the Gell-Mann\textendash Nishijima
 relation

\begin{equation}
Q=I_{3}+\frac{Y}{2}.\label{eq:18}
\end{equation}

\subsection{Flavor operators}

\noindent From the elements of $\mathcal{S}$ it is possible to obtain
four additional groups that commute with $\textrm{SU}(2)_{L}\otimes\text{U}(1)_{Y}$, two $\textrm{SU}(2)$s  and two $\textrm{U}(1)$s.
These groups  
  give zero eigenvalue when acting on non-fermionic states.
Thus, they are suitably interpreted as flavor groups, denoted by $\textrm{SU}(2)_{f}$,
$\textrm{SU}(2)_{\hat{f}}$, $\textrm{U}(1)_{f}$, and $\textrm{U}(1)_{\hat{f}}$.
Their generators are

\begin{equation}
\begin{split}f_{1} & =\frac{i}{8}\left(1+\tilde{\gamma}_{5}\right)\left(1+i\gamma^{5}\gamma^{6}\right)\gamma^{7},\\
f_{2} & =\frac{i}{8}\left(1+\tilde{\gamma}_{5}\right)\left(1+i\gamma^{5}\gamma^{6}\right)\gamma^{8},\\
f_{3} & =\frac{i}{8}\left(1+\tilde{\gamma}_{5}\right)\left(1+i\gamma^{5}\gamma^{6}\right)\gamma^{7}\gamma^{8},
\end{split}
\label{eq:19}
\end{equation}

\begin{equation}
\begin{split}\hat{f}_{1} & =\frac{i}{8}\left(1-\tilde{\gamma}_{5}\right)\left(1+i\gamma^{5}\gamma^{6}\right)\gamma^{7},\\
\hat{f}_{2} & =\frac{i}{8}\left(1-\tilde{\gamma}_{5}\right)\left(1+i\gamma^{5}\gamma^{6}\right)\gamma^{8},\\
\hat{f}_{3} & =\frac{i}{8}\left(1-\tilde{\gamma}_{5}\right)\left(1+i\gamma^{5}\gamma^{6}\right)\gamma^{7}\gamma^{8},
\end{split}
\label{eq:19-1}
\end{equation}

\noindent respectively for by $\textrm{SU}(2)_{f}$ and $\textrm{SU}(2)_{\hat{f}}$,
and

\begin{equation}
f_{0}=i\gamma^{5}\gamma^{6}\tilde{\gamma}_{5},\label{eq:19-2}
\end{equation}

\begin{equation}
\hat{f}_{0}=i\gamma^{5}\gamma^{6},\label{eq:19-3}
\end{equation}

\noindent for $\textrm{U}(1)_{f}$, and $\textrm{U}(1)_{\hat{f}}$.
In terms of the projectors in Eq. (\ref{eq:14-1}) the decomposition
of the diagonal generators $f_{3}$ and $\hat{f}_{3}$ is

\begin{equation}
f_{3}=\frac{1}{2}\left(P_{R1}-P_{R2}\right),\label{eq:37}
\end{equation}

\begin{equation}
\hat{f}_{3}=\frac{1}{2}\left(P_{L1}-P_{L2}\right).\label{eq:38}
\end{equation}

The matrix space is schematically represented in Fig.4, with the operators that classify the states along the diagonal, and the states, to be described below, appearing as off-diagonal matrix elements.

\subsection{Scalar sector}

We obtain
operators that transform as Lorentz scalars from elements of $\mathcal{S}$ multiplied by $\gamma_{0}$\footnote{This property depends also on the    operator  structure, and not
only on multiplication by $\gamma_{0}$, so the latter is a necessary
but not sufficient condition.}. Within such a set, of the     
form $S\gamma_{0}$, with $S\in\mathcal{S}$, in a Hamiltonian description, we concentrate on  mass operators  
which commute with the charge operator.
There are eight of these operators, shown in Table I, with a maximal commuting
set, among them, given by either $M_{1}$ to $M_{4}$ or $M_{5}$
to $M_{8}$. 
\noindent 
\begin{table}[t]
\begin{centering}
\begin{tabular}{ll}
\toprule 
$M_{1}=\gamma^0$ & $M_{5}=i\tilde{\gamma}_{5}\gamma^0$\tabularnewline
\midrule 
$M_{2}=i\gamma^{5}\gamma^{6}\gamma^0$ & $M_{6}=\gamma^{5}\gamma^{6}\tilde{\gamma}_{5}\gamma^0$\tabularnewline
\midrule 
$M_{3}=i\gamma^{7}\gamma^{8}\gamma^0$ & $M_{7}=\gamma^{7}\gamma^{8}\tilde{\gamma}_{5}\gamma^0$\tabularnewline
\midrule 
$M_{4}=\gamma^{5}\gamma^{6}\gamma^{7}\gamma^{8}\gamma^0$ & $M_{8}=i\gamma^{5}\gamma^{6}\gamma^{7}\gamma^{8}\tilde{\gamma}_{5}\gamma^0$\tabularnewline
\bottomrule
\end{tabular}
\par\end{centering}
\caption{Hamiltonian mass operators of the form $S\gamma_{0}$, with $S\in\mathcal{S}$.
Operators $M_{1}$ to $M_{4}$ form a maximal commuting set, and
so do $M_{5}$ to $M_{8}$. The two sets are connected by a transformation
involving $\tilde{\gamma}_{5}$. }
\end{table}

We look among the set   for those with the  Higgs-field quantum
numbers: $\textrm{SU}(2)_{L}$ doublets, and hypercharge $\pm1$, when classified by the relevant $\textrm{SU}(2)_{L}\otimes\text{U}(1)_{Y}$
operators following Eq. (\ref{eq:6}), and find that there are at most
two such doublets. This can be seen in Fig. 4, where the chiral projections
of the diagonal operators $B$, $I_{3}$ and $Y$, according to Eqs.
(\ref{eq:31}), (\ref{eq:29}), (\ref{eq:33}),  and Fig. 3, are grouped
together in the sets ${\bf Q}_{R}=\frac{1}{2}\left(1+\tilde{\gamma}_{5}\right)\left(B,I_{3},Y\right)$
and ${\bf Q}_{L}=\frac{1}{2}\left(1-\tilde{\gamma}_{5}\right)\left(B,I_{3},Y\right)$.
The off-diagonal matrix elements represent the states, and following
the matrix product rule, they are multiplied from the left by diagonal
operators in the same row, and from the right by diagonal operators
in the same column, producing the eigenvalues corresponding to the
weights in Eqs. (\ref{eq:31}), (\ref{eq:29}), and  (\ref{eq:33}).

Let us consider the matrix elements in rows 13,  14, and columns
7, 8 in Fig. 4, labeled $\varphi_{L,1,2}^{+}$, and belonging to the left-chiral
projection of the matrix space. Their  classification stems from action from the left   
by ${\bf Q}_{L}$, and from the right by ${\bf Q}_{R}$; hence Eq. (\ref{eq:6})
gives the eigenvalues $\left(1/3,1/2,1/3\right)-\left(1/3,0,-2/3\right)=\left(0,1/2,1\right)$,
for baryon number, isospin and hypercharge, respectively. Thus, they
have the quantum numbers of an $\textrm{SU}(2)_{L}$ upper
  doublet-component  charged scalar field. Now  the matrix elements
in rows 5, 6, and columns 15, 16, labeled $\varphi_{R,1,2}^{+}$, belong
to the right-chiral projection of the matrix space; they are classified
from the left by ${\bf Q}_{R}$, and from the right by ${\bf Q}_{L}$, giving
the same quantum numbers as the previous ones because $\left(1/3,0,4/3\right)-\left(1/3,-1/2,1/3\right)=\left(0,1/2,1\right)$.
Therefore, a SM Higgs field (non-chirally projected) must be given
by a combination of them. With the eight real degrees of freedom of
the two blocks we obtain 4 complex elements, from which we can have
two linearly independent, scalar charged fields: $\phi_{1}^{+}$
and $\phi_{2}^{+}$.  An analogous construction produces two neutral
Higgs fields, denoted by $\phi_{1}^{0}$ and $\phi_{2}^{0}$, and
connected to the charged ones by the ladder operators of $\textrm{SU}(2)_{L}$. 
The scalar Higgs doublets are given in Table II.

\noindent 
\begin{table}[t]
\begin{centering}
\begin{tabular}{cccc}
\toprule 
Baryon number zero, Higgs-like scalars & $I_{3}$ & $Y$ & $Q$\tabularnewline
\midrule
\midrule 
$\boldsymbol{\phi}_{1}=\begin{pmatrix}\phi_{1}^{+}\\
\phi_{1}^{0}
\end{pmatrix}=\begin{pmatrix}\frac{1}{8}\left(1-i\gamma^{5}\gamma^{6}\right)\left(\gamma^{7}+i\gamma^{8}\right)\gamma_{0}\\
\frac{1}{8}\left(1-i\gamma^{5}\gamma^{6}\right)\left(1+i\gamma^{7}\gamma^{8}\tilde{\gamma}_{5}\right)\gamma_{0}
\end{pmatrix}$ & $\begin{array}{r}
1/2\\
-1/2
\end{array}$ & $1$ & $\begin{array}{r}
1\\
0
\end{array}$\tabularnewline
\midrule 
$\boldsymbol{\phi}_{2}=\begin{pmatrix}\phi_{2}^{+}\\
\phi_{2}^{0}
\end{pmatrix}=\begin{pmatrix}\frac{1}{8}\left(1-i\gamma^{5}\gamma^{6}\right)\left(\gamma^{7}+i\gamma^{8}\right)\tilde{\gamma}_{5}\gamma_{0}\\
\frac{i}{8}\left(1-i\gamma^{5}\gamma^{6}\right)\left(1+i\gamma^{7}\gamma^{8}\tilde{\gamma}_{5}\right)\gamma^{7}\gamma^{8}\gamma_{0}
\end{pmatrix}$ & $\begin{array}{r}
1/2\\
-1/2
\end{array}$ & $1$ & $\begin{array}{r}
1\\
0
\end{array}$\tabularnewline
\bottomrule
\end{tabular}
\par\end{centering}
\caption{Scalar Higgs-like doublets}
\end{table}

Only some combinations of mass operators correspond to the neutral Higgs fields, so that the latter can be given in terms of the former, e.g.,

\begin{equation}
\begin{split}\phi_{1}^{0}=\frac{1}{8}\left(M_{1}-M_{2}\right)+\frac{i}{8}\left(M_{7}-M_{8}\right),\\
\phi_{2}^{0}=\frac{1}{8}\left(M_{3}+M_{4}\right)-\frac{i}{8}\left(M_{5}+M_{6}\right).
\end{split}
\label{eq:39.1}
\end{equation}

\noindent There are also mass-operator terms that do not correspond to any combination of them. Thus, after electroweak symmetry breaking when only the neutral Higgs
fields survive, we consider the following linear combinations of the
neutral Higgs fields and their Hermitian conjugates, and of the Hamiltonian
mass operators, taken from the first set in Table I

\begin{align}
\begin{split}\mathcal{M}_{1}= & a_{1}\left(\phi_{1}^{0}+\phi_{1}^{0\dagger}\right)+b_{1}\left(\phi_{2}^{0}+\phi_{2}^{0\dagger}\right),\\
\mathcal{M}_{2}= & \frac{a_{2}}{4}\left(M_{1}+M_{2}\right)+\frac{b_{2}}{4}\left(M_{3}-M_{4}\right),
\end{split}
\label{eq:39}
\end{align}

\noindent with $a_{1,2}$ and $b_{1,2}$ real parameters, and $\mathcal{M}_{2}$ cannot be given in terms of the neutral Higgses (the $1/4$ factor is just for normalization convenience). In Ref. \citep{Besprosvany:2014lwa} a Lagrangian for the
Higgs-field doublets was given for a reduced model with only two quark
flavors. As for the $\mathcal{M}_{2}$ operator in Eq. (\ref{eq:39}),  
it is a non Higgs-like  scalar that also classifies the massive
quark states, fulfilling the same function t as a flavon field\citep{FROGGATT1979277,Bauer2015}. In fact, $\mathcal{M}_{2}$  produces a horizontal mass hierarchy, as is shown in Subsection III.E.

\subsection{Massless quarks }

\begin{figure}[t]
\begin{centering}
\begin{tikzpicture}[scale=1,every node/.style={scale=0.7}] 
\draw (0,0) grid (8,8); 


\draw (1,6.5) -- (2,6.5);
\draw (1.5,6) -- (1.5,7);
\draw (3,4.5) -- (4,4.5);
\draw (3.5,4) -- (3.5,5);
\draw (5,2.5) -- (6,2.5);
\draw (5.5,2) -- (5.5,3);
\draw (7,0.5) -- (8,0.5);
\draw (7.5,0) -- (7.5,1);
\draw (0,7.5) -- (1,7.5);
\draw (0.5,7) -- (0.5,8);
\draw (2,5.5) -- (3,5.5);
\draw (2.5,5) -- (2.5,6);
\draw (4,3.5) -- (5,3.5);
\draw (4.5,3) -- (4.5,4);
\draw (6,1.5) -- (7,1.5);
\draw (6.5,1) -- (6.5,2);

\draw [fill=magenta,thick] (0,7.5) rectangle (0.5,8);
\draw [fill=magenta,thick] (0.5,7) rectangle (1,7.5);
\draw [fill=magenta,thick] (1,6.5) rectangle (1.5,7);
\draw [fill=magenta,thick] (1.5,6) rectangle (2,6.5);
\draw [fill=yellow,thick] (2,5.5) rectangle (2.5,6);
\draw [fill=yellow,thick] (2.5,5) rectangle (3,5.5);
\draw [fill=yellow,thick] (3,4.5) rectangle (3.5,5);
\draw [fill=yellow,thick] (3.5,4) rectangle (4,4.5);

\draw [fill=pink,thick] (4,3.5) rectangle (4.5,4);
\draw [fill=pink,thick] (4.5,3) rectangle (5,3.5);
\draw [fill=pink,thick] (5,2.5) rectangle (5.5,3);
\draw [fill=pink,thick] (5.5,2) rectangle (6,2.5);
\draw [fill=lime,thick] (6,1.5) rectangle (6.5,2);
\draw [fill=lime,thick] (6.5,1) rectangle (7,1.5);
\draw [fill=lime,thick] (7,0.5) rectangle (7.5,1);
\draw [fill=lime,thick] (7.5,0) rectangle (8,0.5);

\draw [fill=green,thick] (0,5) rectangle (1,6);
\draw [fill=green,thick] (1,5) rectangle (2,6);
\draw [fill=green,thick] (4,5) rectangle (5,6);
\draw [fill=green,thick] (5,5) rectangle (6,6);
\draw [fill=cyan,thick] (0,4) rectangle (1,5);
\draw [fill=cyan,thick] (1,4) rectangle (2,5);
\draw [fill=cyan,thick] (4,4) rectangle (5,5);
\draw [fill=cyan,thick] (5,4) rectangle (6,5);

\draw [fill=orange,thick] (0,1) rectangle (1,2);
\draw [fill=orange,thick] (1,1) rectangle (2,2);
\draw [fill=orange,thick] (4,1) rectangle (5,2);
\draw [fill=orange,thick] (5,1) rectangle (6,2);
\draw [fill=brown,thick] (0,0) rectangle (1,1);
\draw [fill=brown,thick] (1,0) rectangle (2,1);
\draw [fill=brown,thick] (4,0) rectangle (5,1);
\draw [fill=brown,thick] (5,0) rectangle (6,1);

\draw [fill=teal,thick] (7,4) rectangle (8,5);
\draw [fill=teal,thick] (7,5) rectangle (8,6);
\draw [fill=teal,thick] (3,0) rectangle (4,1);
\draw [fill=teal,thick] (3,1) rectangle (4,2);



\node at (0.25,7.75) {$\mathbf{f_{3}}$};
\node at (0.75,7.25) {$\mathbf{f_{3}}$};
\node at (1.25,6.75) {$\mathbf{f_{3}}$}; 
\node at (1.75,6.25) {$\mathbf{f_{3}}$}; 
\node at (4.25,3.75) {$\mathbf{\hat{f}_{3}}$};
\node at (4.75,3.25) {$\mathbf{\hat{f}_{3}}$};
\node at (5.25,2.75) {$\mathbf{\hat{f}_{3}}$};
\node at (5.75,2.25) {$\mathbf{\hat{f}_{3}}$};

\node at (2.25,5.75) {$\mathbf{Q_R}$};
\node at (2.75,5.25) {$\mathbf{Q_R}$};
\node at (3.25,4.75) {$\mathbf{Q_R}$};
\node at (3.75,4.25) {$\mathbf{Q_R}$};
\node at (6.25,1.75) {$\mathbf{Q_L}$};
\node at (6.75,1.25) {$\mathbf{Q_L}$};
\node at (7.25,0.75) {$\mathbf{Q_L}$};
\node at (7.75,0.25) {$\mathbf{Q_L}$};


\node at (3.5,1.5) {$\mathbf{\varphi^{+}_{L,1,2}}$};
\node at (3.5,0.5) {$\mathbf{\varphi^{0}_{L,1,2}}$};
\node at (7.5,5.5) {$\mathbf{\varphi^{+}_{R,1,2}}$};
\node at (7.5,4.5) {$\mathbf{\varphi^{0}_{R,1,2}}$};


\node at (0.5,5.5) {$\mathbf{U_{R,1}^1}$};
\node at (0.5,4.5) {$\mathbf{D_{R,1}^1}$};
\node at (1.5,5.5) {$\mathbf{U_{R,2}^1}$};
\node at (1.5,4.5) {$\mathbf{D_{R,2}^1}$};
\node at (4.5,5.5) {$\mathbf{U_{R,3}^1}$};
\node at (4.5,4.5) {$\mathbf{D_{R,3}^1}$};
\node at (5.5,5.5) {$\mathbf{U_{R,4}^1}$};
\node at (5.5,4.5) {$\mathbf{D_{R,4}^1}$};

\node at (0.5,1.5) {$\mathbf{U_{L,1}^1}$};
\node at (0.5,0.5) {$\mathbf{D_{L,1}^1}$};
\node at (1.5,1.5) {$\mathbf{U_{L,2}^1}$};
\node at (1.5,0.5) {$\mathbf{D_{L,2}^1}$};
\node at (4.5,1.5) {$\mathbf{U_{L,3}^1}$};
\node at (4.5,0.5) {$\mathbf{D_{L,3}^1}$};
\node at (5.5,1.5) {$\mathbf{U_{L,4}^1}$};
\node at (5.5,0.5) {$\mathbf{D_{L,4}^1}$};


\foreach \i/\valorI in {1/1,2/3,3/5,4/7,5/9,6/11,7/13,8/15}
\foreach \j/\valorP in {1/2,2/4,3/6,4/8,5/10,6/12,7/14,8/16}  
{     
\node[anchor=south] at (\i-0.75,8) {$\valorI$}; 
\node[anchor=south] at (\j-0.25,8) {$\valorP$};    
\node[anchor=east] at (0,-\i+8.75) {$\valorI$};
\node[anchor=east] at (0,-\j+8.25) {$\valorP$};
\node[anchor=south] at (\i-0.75,-0.4) {$\valorI$}; 
\node[anchor=south] at (\j-0.25,-0.4) {$\valorP$};   
\node[anchor=east] at (8.4,-\i+8.75) {$\valorI$};
\node[anchor=east] at (8.4,-\j+8.25) {$\valorP$};
} 
\end{tikzpicture}
\par\end{centering}
\caption{Matrix representation of operators, massless quarks $\left(U_{L,R}^{i},\,D_{L,R}^{i},\,i=1,\ldots,4\right)$
and Higgs $\left(\phi_{1,2}^{+},\,\phi_{1,2}^{0}\right)$ degrees
of freedom in $\left(7+1\right)$-d spin space. The chiral projections
of the diagonal operators $B$, $I_{3}$ and $Y$ are grouped  
and represented by the sets ${\bf Q}_{R}=\frac{1}{2}\left(1+\tilde{\gamma}_{5}\right)\left(B,I_{3},Y\right)$
and ${\bf Q}_{L}=\frac{1}{2}\left(1-\tilde{\gamma}_{5}\right)\left(B,I_{3},Y\right)$.
Following matrix multiplication rules, operators act from the left
on states in the same row, and from the right on states in the same
column.}
\end{figure}
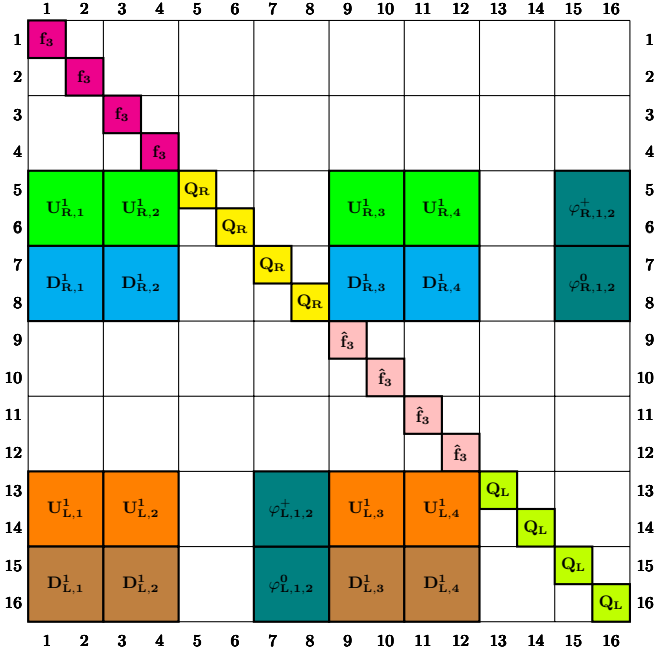

The   massless-fermion spectrum  consists of states with the structure
of Eq. (\ref{eq:10}), and with the corresponding quantum numbers,  
when classified by baryon number, isospin, and hypercharge. The dimensionality of the
matrix space gives rise to four generations of massless quarks\footnote{The space dimension is not big enough to accommodate color SU(3) of
QCD, but this can be done in 9+1 dimensions, see Ref. \citep{Besprosvany:2002tv}.}. In Fig. 4, the elements labeled $U_{L}^{i}$, $i=1,\ldots,4$ in
rows 13 and 14 represent all possible degrees of freedom with the
quantum numbers $\left(1/3,1/2,1/3\right)$ for the latter operators, respectively, and the corresponding flavor. Thus,
they are interpreted as left-handed quarks of the up-type. There are
4 degrees of freedom for each $U_{L}^{i}$  accounting for the two
possible spin directions (up/down) and chirality (left/right-handed),
so in this respect, even though they are represented by $16\times16$
matrices, fermions in the extended spin space carry the same degrees
of freedom as a massless  4-d  Weyl spinor. This latter feature is
a consequence of maintaining Lorentz symmetry in 4-d.

Analogous results hold for right-handed up-type quarks, shown in rows
5 and 6 in Fig. 4, and for down-type quarks, both left-handed and
right-handed, shown respectively in rows 15 and 16, and rows 7 and
8 in Fig. 4. Thus, there are in total eight left-handed quarks arranged
in four $\textrm{SU}(2)_{L}$ doublets, with elements connected vertically
by the ladder operators $\frac{1}{\sqrt{2}}\left(I_{1}\pm iI_{2}\right)$,
and eight right-handed $\textrm{SU}(2)_{L}$ singlets. The quark states
are given in Tables III and IV, together with their quantum numbers.

The flavor operators provide further symmetries:
members of the first two families with the same charge are horizontal
$\textrm{SU}(2)_{f1}$ doublets when classified with the ladder operators
$\frac{1}{\sqrt{2}}\left(f_{1}\pm if_{2}\right)$, and the same is
true for the last two families when classified with $\frac{1}{\sqrt{2}}\left(\hat{f}_{1}\pm i\hat{f}_{2}\right)$
of $\textrm{SU}(2)_{f2}$, e.g.

\begin{equation}
\begin{array}{ccc}
U_{L,1}^{1}\frac{1}{\sqrt{2}}\left(f_{1}-if_{2}\right)=\frac{i}{\sqrt{2}}U_{L,2}^{1}, &  & U_{L,2}^{1}\frac{1}{\sqrt{2}}\left(f_{1}-if_{2}\right)=0,\\
\\
U_{L,2}^{1}\frac{1}{\sqrt{2}}\left(f_{1}+if_{2}\right)=-\frac{i}{\sqrt{2}}U_{L,1}^{1}, &  & U_{L,1}^{1}\frac{1}{\sqrt{2}}\left(f_{1}+if_{2}\right)=0,
\end{array}\label{eq:39-1}
\end{equation}

\noindent where, in accord with the discussion below (\ref{eq:12-1}), flavor operators act non-trivially only from the rhs. This horizontal flavor symmetry also applies to right-handed quarks. 

There is also the possibility of projecting out a generation. As an example, let us consider the operator

\begin{equation}
F=-\frac{1}{4}\left(\hat{f}_{0}-4\hat{f}_{3}-8f_{3}\right),\label{eq:37-1}
\end{equation}

\noindent which yields $\left[F,U_{L,i}^{1}\right]=\lambda_{i}U_{L,i}^{1}$,
$\left[F,D_{L,i}^{1}\right]=\lambda_{i}D_{L,i}^{1}$, $\left[F,U_{R,i}^{1}\right]=\lambda_{i}U_{R,i}^{1}$,
$\left[F,D_{R,i}^{1}\right]=\lambda_{i}D_{R,i}^{1}$, $i=1,\ldots,4$,
with $\lambda_{1}=3/2$, $\lambda_{2}=-1/2$, $\lambda_{3}=1$, and $\lambda_{4}=0$.
Therefore, it provides a horizontal (family) $\text{U}(1)$ symmetry
for the quark states, with the zero value for $\lambda_{4}$ interpreted
as projecting the fourth quark family out of flavor space, so effectively,
we get three massless generations. This operator commutes with $B$, $Q$ and $\mathcal{M}_1$, respectively in Eqs. (\ref{eq:29}), (\ref{eq:18}), and (\ref{eq:39}), but not with $\mathcal{M}_2$ in the latter. Its complete characterization requires further investigation, which is left for future work and at present it will be no longer considered. 

\begin{table}[t]
\begin{centering}
\begin{tabular}{>{\raggedright}p{0.55\textwidth}c>{\centering}p{0.06\textwidth}>{\centering}p{0.06\textwidth}>{\centering}p{0.06\textwidth}>{\centering}p{0.06\textwidth}}
\toprule 
Baryon number 1/3, hypercharge 1/3, and polarization 1/2

$\left(\text{operator }\frac{3}{2}iB\gamma^{1}\gamma^{2}\right)$,
left-handed quark doublets & $I_{3}$ & $Q$ & $f_{3}$ & $\hat{f}_{3}$ & $F$\tabularnewline
\midrule
\midrule 
$\mathbf{Q}^{1}_{L,1}=\begin{pmatrix}U^{1}_{L,1}\\
D^{1}_{L,1}
\end{pmatrix}=\begin{pmatrix}\frac{1}{16}\left(1-\tilde{\gamma}_{5}\right)\left(\gamma^{5}-i\gamma^{6}\right)\left(\gamma^{7}+i\gamma^{8}\right)\left(\gamma^{0}+\gamma^{3}\right)\\
\frac{1}{16}\left(1-\tilde{\gamma}_{5}\right)\left(\gamma^{5}-i\gamma^{6}\right)\left(1-i\gamma^{7}\gamma^{8}\right)\left(\gamma^{0}+\gamma^{3}\right)
\end{pmatrix}$ & $\begin{array}{r}
1/2\\
-1/2
\end{array}$ & $\begin{array}{r}
2/3\\
-1/3
\end{array}$ & $\begin{array}{r}
-1/2\\
-1/2
\end{array}$ & $0$ & $\begin{array}{r}
3/2\\
3/2
\end{array}$\tabularnewline
\midrule 
$\mathbf{Q}^{1}_{L,2}=\begin{pmatrix}U^{1}_{L,2}\\
D^{1}_{L,2}
\end{pmatrix}=\begin{pmatrix}\frac{1}{16}\left(1-\tilde{\gamma}_{5}\right)\left(\gamma^{5}-i\gamma^{6}\right)\left(1+i\gamma^{7}\gamma^{8}\right)\left(\gamma^{0}+\gamma^{3}\right)\\
\frac{1}{16}\left(1-\tilde{\gamma}_{5}\right)\left(\gamma^{5}-i\gamma^{6}\right)\left(\gamma^{7}-i\gamma^{8}\right)\left(\gamma^{0}+\gamma^{3}\right)
\end{pmatrix}$ & $\begin{array}{r}
1/2\\
-1/2
\end{array}$ & $\begin{array}{r}
2/3\\
-1/3
\end{array}$ & $\begin{array}{r}
1/2\\
1/2
\end{array}$ & $0$ & $\begin{array}{r}
-1/2\\
-1/2
\end{array}$\tabularnewline
\midrule 
$\mathbf{Q}^{1}_{L,3}=\begin{pmatrix}U^{1}_{L,3}\\
D^{1}_{L,3}
\end{pmatrix}=\begin{pmatrix}\frac{1}{16}\left(1-\tilde{\gamma}_{5}\right)\left(\gamma^{5}-i\gamma^{6}\right)\left(\gamma^{7}+i\gamma^{8}\right)\gamma^{0}\left(\gamma^{0}-\gamma^{3}\right)\\
\frac{1}{16}\left(1-\tilde{\gamma}_{5}\right)\left(\gamma^{5}-i\gamma^{6}\right)\left(1-i\gamma^{7}\gamma^{8}\right)\gamma^{0}\left(\gamma^{0}-\gamma^{3}\right)
\end{pmatrix}$ & $\begin{array}{r}
1/2\\
-1/2
\end{array}$ & $\begin{array}{r}
2/3\\
-1/3
\end{array}$ & $0$ & $\begin{array}{r}
-1/2\\
-1/2
\end{array}$ & $\begin{array}{r}
1\\
1
\end{array}$\tabularnewline
\midrule 
$\mathbf{Q}^{1}_{L,4}=\begin{pmatrix}U^{1}_{L,4}\\
D^{1}_{L,4}
\end{pmatrix}=\begin{pmatrix}\frac{1}{16}\left(1-\tilde{\gamma}_{5}\right)\left(\gamma^{5}-i\gamma^{6}\right)\left(1+i\gamma^{7}\gamma^{8}\right)\gamma^{0}\left(\gamma^{0}-\gamma^{3}\right)\\
\frac{1}{16}\left(1-\tilde{\gamma}_{5}\right)\left(\gamma^{5}-i\gamma^{6}\right)\left(\gamma^{7}-i\gamma^{8}\right)\gamma^{0}\left(\gamma^{0}-\gamma^{3}\right)
\end{pmatrix}$ & $\begin{array}{r}
1/2\\
-1/2
\end{array}$ & $\begin{array}{r}
2/3\\
-1/3
\end{array}$ & $0$ & $\begin{array}{r}
1/2\\
1/2
\end{array}$ & $0$\tabularnewline
\bottomrule
\end{tabular}
\par\end{centering}
\caption{Massless left-handed quark weak isospin doublets. Gauge and Lorentz
operators act from the left and trivially from the right, while the
reverse is true for flavor operators. To obtain the -1/2 polarization, the
replacement must be made $\left(\gamma^{0}+\gamma^{3}\right)\rightarrow\left(\gamma^{1}-i\gamma^{2}\right)$, for  $\mathbf{Q}^{1}_{L,1}$, $\mathbf{Q}^{1}_{L,1}$, and 
 $\left(\gamma^{0}-\gamma^{3}\right)\rightarrow\left(\gamma^{1}-i\gamma^{2}\right)$,
 for  $\mathbf{Q}^{1}_{L,3}$, $\mathbf{Q}^{4}_{L,4}$. } 
\end{table}

\begin{table}[t]
\begin{centering}
\begin{tabular}{>{\raggedright}p{0.55\textwidth}ccccc}
\toprule 
Baryon number 1/3, hypercharge 4/3 and -2/3, and polarization 1/2

$\left(\text{operator }\frac{3}{2}iB\gamma^{1}\gamma^{2}\right)$,
right-handed quark singlets & $Y$ & $Q$ & $f_{3}$ & $\hat{f}_{3}$ & $F$\tabularnewline
\midrule
\midrule 
$\begin{array}{l}
U^{1}_{R,1}=\frac{1}{16}\left(1+\tilde{\gamma}_{5}\right)\left(\gamma^{5}-i\gamma^{6}\right)\left(\gamma^{7}+i\gamma^{8}\right)\gamma^{0}\left(\gamma^{0}+\gamma^{3}\right)\\
D^{1}_{R,1}=\frac{1}{16}\left(1+\tilde{\gamma}_{5}\right)\left(\gamma^{5}-i\gamma^{6}\right)\left(1-i\gamma^{7}\gamma^{8}\right)\gamma^{0}\left(\gamma^{0}+\gamma^{3}\right)
\end{array}$ & $\begin{array}{r}
4/3\\
-2/3
\end{array}$ & $\begin{array}{r}
2/3\\
-1/3
\end{array}$ & $\begin{array}{r}
-1/2\\
-1/2
\end{array}$ & $0$ & $\begin{array}{r}
3/2\\
3/2
\end{array}$\tabularnewline
\midrule 
$\begin{array}{l}
U^{1}_{R,2}=\frac{1}{16}\left(1+\tilde{\gamma}_{5}\right)\left(\gamma^{5}-i\gamma^{6}\right)\left(1+i\gamma^{7}\gamma^{8}\right)\gamma^{0}\left(\gamma^{0}+\gamma^{3}\right)\\
D^{1}_{R,2}=\frac{1}{16}\left(1+\tilde{\gamma}_{5}\right)\left(\gamma^{5}-i\gamma^{6}\right)\left(\gamma^{7}-i\gamma^{8}\right)\gamma^{0}\left(\gamma^{0}+\gamma^{3}\right)
\end{array}$ & $\begin{array}{r}
4/3\\
-2/3
\end{array}$ & $\begin{array}{r}
2/3\\
-1/3
\end{array}$ & $\begin{array}{r}
1/2\\
1/2
\end{array}$ & $0$ & $\begin{array}{r}
-1/2\\
-1/2
\end{array}$\tabularnewline
\midrule 
$\begin{array}{l}
U^{1}_{R,3}=\frac{1}{16}\left(1+\tilde{\gamma}_{5}\right)\left(\gamma^{5}-i\gamma^{6}\right)\left(\gamma^{7}+i\gamma^{8}\right)\left(\gamma^{0}-\gamma^{3}\right)\\
D^{1}_{R,3}=\frac{1}{16}\left(1+\tilde{\gamma}_{5}\right)\left(\gamma^{5}-i\gamma^{6}\right)\left(1-i\gamma^{7}\gamma^{8}\right)\left(\gamma^{0}-\gamma^{3}\right)
\end{array}$ & $\begin{array}{r}
4/3\\
-2/3
\end{array}$ & $\begin{array}{r}
2/3\\
-1/3
\end{array}$ & $0$ & $\begin{array}{r}
-1/2\\
-1/2
\end{array}$ & $\begin{array}{r}
1\\
1
\end{array}$\tabularnewline
\midrule 
$\begin{array}{l}
U^{1}_{R,4}=\frac{1}{16}\left(1+\tilde{\gamma}_{5}\right)\left(\gamma^{5}-i\gamma^{6}\right)\left(1+i\gamma^{7}\gamma^{8}\right)\left(\gamma^{0}-\gamma^{3}\right)\\
D^{1}_{R,4}=\frac{1}{16}\left(1+\tilde{\gamma}_{5}\right)\left(\gamma^{5}-i\gamma^{6}\right)\left(\gamma^{7}-i\gamma^{8}\right)\left(\gamma^{0}-\gamma^{3}\right)
\end{array}$ & $\begin{array}{r}
4/3\\
-2/3
\end{array}$ & $\begin{array}{r}
2/3\\
-1/3
\end{array}$ & $0$ & $\begin{array}{r}
1/2\\
1/2
\end{array}$ & $0$\tabularnewline
\bottomrule
\end{tabular}
\par\end{centering}
\caption{Massless right-handed quark weak isospin singlets. Gauge and Lorentz
operators act from the left and trivially from the right, while the
reverse is true for flavor operators. To obtain the -1/2 polarization, the
replacement must be made $\left(\gamma^{0}+\gamma^{3}\right)\rightarrow\left(\gamma^{1}-i\gamma^{2}\right)$, for  $U^{1}_{R,1}$, $U^{1}_{R,2}$, $D^{1}_{R,1}$, $D^{1}_{R,2}$, and 
 $\left(\gamma^{0}-\gamma^{3}\right)\rightarrow\left(\gamma^{1}-i\gamma^{2}\right)$,
 for  $U^{1}_{R,3}$, $U^{1}_{R,4}$, $D^{1}_{R,3}$, $D^{1}_{R,4}$.}
\end{table}

\subsection{Mass hierarchy effects}

Let us consider the general mass operator

\begin{equation}
\Omega\equiv\mathcal{M}_{1}+\mathcal{M}_{2},\label{eq:43}
\end{equation}

\noindent with $\mathcal{M}_{1}$ and $\mathcal{M}_{2}$ given in
Eq. (\ref{eq:39}). This operator can be diagonalized with the massless
quark states to obtain massive ones, e. g., the combinations

\begin{align}
\begin{split}U_{M}= & \frac{1}{\sqrt{2}}\left(U_{L,1}^{1}+U_{R,1}^{1}\right),\\
D_{M}= & \frac{1}{\sqrt{2}}\left(D_{L,1}^{1}-D_{R,1}^{1}\right),
\end{split}
\label{eq:43-2}
\end{align}

\noindent have the quantum numbers of massive quarks of the up and
down-type, respectively. Considering only the $\mathcal{M}_{1}$ part
in Eq. (\ref{eq:43}), that is setting $a_{2}=b_{2}=0$, we obtain
the eigenvalues

\begin{gather}
\begin{gathered}\dfrac{a_{1}+b_{1}}{2},\\
\dfrac{a_{1}-b_{1}}{2},
\end{gathered}
\label{eq:42}
\end{gather}

\noindent respectively for $U_{M}$ and $D_{M}$ in Eq. (\ref{eq:43-2}), 
which constitute a vertical mass hierarchy depending on the order
of the parameters involved, with the identifications

\begin{align}
\begin{split}a_{1}= & m_{u}+m_{d},\\
b_{1}= & m_{u}-m_{d},
\end{split}
\label{eq:42-1}
\end{align}

\noindent where $m_{u}$ and $m_{d}$ are current quark masses of
the up and down type, respectively, for a given family. However, the structure of the states in Eq. (\ref{eq:43-2}) does not allow them to be eigenstates of the $\mathcal{M2}$
operator in Eq. (\ref{eq:43}), responsible for the horizontal mass hierarchy effect as we next show. In order for them to be also eigenstates of $\mathcal{M2}$, and therefore  diagonalize $\Omega$, another generation of massless quarks, in a left and right combination, must be added, and the relative phases fixed, so that the states in Eq. (\ref{eq:43-2}) change to

\begin{align}
\begin{split}U_{M,1}^{1}= & \frac{1}{2}\left(-U_{L,1}^{1}-U_{R,1}^{1}+U_{L,3}^{1}+U_{R,3}^{1}\right)\\
D_{M,1}^{1}= & \dfrac{1}{2}\left(-D_{L,1}^{1}+D_{R,1}^{1}+D_{L,3}^{1}-D_{R,3}^{1}\right)
\end{split}
\label{eq:41-1}
\end{align}

The four generations of massive quarks is shown in Table V, and the vertical hierarchy effect in Eq. (\ref{eq:42}) remains valid for $U_{M,i}^{1}$ and $D_{M,i}^{1}$ quarks, $i=1,\ldots,4$,

\noindent 
\begin{table}[t]
\noindent \begin{centering}
\begin{tabular}{lrcrc}
\toprule 
\multicolumn{1}{>{\raggedright}p{6cm}}{Baryon number 1/3 and polarization 1/2 $\left(\text{operator }\frac{3}{2}iB\gamma^{1}\gamma^{2}\right)$,
massive quarks} & $Q$ & $\mathcal{M}_{1}$ & $\mathcal{M}_{2}$ & $\Omega$\tabularnewline
\midrule
\midrule 
$U_{M,1}^{1}=\frac{1}{2}\left(-U_{L,1}^{1}-U_{R,1}^{1}+U_{L,3}^{1}+U_{R,3}^{1}\right)$ & $2/3$ & $\frac{a_{1}+b_{1}}{2}$ & $\frac{a_{2}-b_{2}}{2}$ & $\frac{a_{1}+b_{1}+a_{2}-b_{2}}{2}$\tabularnewline
\midrule 
$U_{M,2}^{1}=\frac{1}{2}\left(U_{L,2}^{1}-U_{R,2}^{1}-U_{L,4}^{1}+U_{R,4}^{1}\right)$ & $2/3$ & $\frac{a_{1}+b_{1}}{2}$ & $\frac{a_{2}+b_{2}}{2}$ & $\frac{a_{1}+b_{1}+a_{2}+b_{2}}{2}$\tabularnewline
\midrule 
$U_{M,3}^{1}=-\frac{1}{2}\left(U_{L,1}^{1}+U_{R,1}^{1}+U_{L,3}^{1}+U_{R,3}^{1}\right)$ & $2/3$ & $\frac{a_{1}+b_{1}}{2}$ & -$\frac{a_{2}-b_{2}}{2}$ & $\frac{a_{1}+b_{1}-a_{2}+b_{2}}{2}$\tabularnewline
\midrule 
$U_{M,4}^{1}=\frac{1}{2}\left(U_{L,2}^{1}-U_{R,2}^{1}+U_{L,4}^{1}-U_{R,4}^{1}\right)$ & $2/3$ & $\frac{a_{1}+b_{1}}{2}$ & $-\frac{a_{2}+b_{2}}{2}$ & $\frac{a_{1}+b_{1}-a_{2}-b_{2}}{2}$\tabularnewline
\midrule 
$D_{M,1}^{1}=\frac{1}{2}\left(-D_{L,1}^{1}+D_{R,1}^{1}+D_{L,3}^{1}-D_{R,3}^{1}\right)$ & $-1/3$ & $\frac{a_{1}-b_{1}}{2}$ & $\frac{a_{2}-b_{2}}{2}$ & $\frac{a_{1}-b_{1}+a_{2}-b_{2}}{2}$\tabularnewline
\midrule 
$D_{M,2}^{1}=\frac{1}{2}\left(D_{L,2}^{1}-D_{R,2}^{1}-D_{L,4}^{1}+D_{R,4}^{1}\right)$ & $-1/3$ & $\frac{a_{1}-b_{1}}{2}$ & $\frac{a_{2}+b_{2}}{2}$ & $\frac{a_{1}-b_{1}+a_{2}+b_{2}}{2}$\tabularnewline
\midrule 
$D_{M,3}^{1}=\frac{1}{2}\left(-D_{L,1}^{1}+D_{R,1}^{1}-D_{L,3}^{1}+D_{R,3}^{1}\right)$ & $-1/3$ & $\frac{a_{1}-b_{1}}{2}$ & -$\frac{a_{2}-b_{2}}{2}$ & $\frac{a_{1}-b_{1}-a_{2}+b_{2}}{2}$\tabularnewline
\midrule 
$D_{M,4}^{1}=\frac{1}{2}\left(D_{L,2}^{1}+D_{R,2}^{1}-D_{L,4}^{1}+D_{R,4}^{1}\right)$ & $-1/3$ & $\frac{a_{1}-b_{1}}{2}$ & $-\frac{a_{2}+b_{2}}{2}$ & $\frac{a_{1}-b_{1}-a_{2}-b_{2}}{2}$\tabularnewline
\bottomrule
\end{tabular}
\par\end{centering}
\caption{Massive quark states }
\end{table}

Turning now to the $\mathcal{M}_{2}$ part in
Eq. (\ref{eq:43}), with $a_{1}=b_{1}=0$, and considering the same states in
Eq. (\ref{eq:41-1}), together with the second family from Table V

\begin{equation}
\begin{split}U^{1}_{M,2}= & \frac{1}{2}\left(U^{1}_{L,2}-U^{1}_{R,2}-U^{1}_{L,4}+U^{1}_{R,4}\right)\\
D^{1}_{M,2}= & \dfrac{1}{2}\left(D^{1}_{L,2}+D^{1}_{R,2}-D^{1}_{L,4}-D^{1}_{R,4}\right)
\end{split}
\label{eq:43-1}
\end{equation}

\noindent we obtain the eigenvalues

\begin{equation}
\begin{gathered}\dfrac{a_{2}-b_{2}}{2},\\
\dfrac{a_{2}+b_{2}}{2},
\end{gathered}
\label{eq:46}
\end{equation}

\noindent with the first value corresponding to the generation in
Eq. (\ref{eq:41-1}), and the second to the one in Eq. (\ref{eq:43-1}).
Thus, we see that the operator $\mathcal{M}_{2}$ produces a horizontal
(family) mass hierarchy. The massive quark states are shown in Table
V, with the eigenvalues from operators in Eqs. (\ref{eq:39}) and
(\ref{eq:43}).

\section{Concluding remarks}

In this paper, we  presented an electroweak quark model derived from  an extended (7+1)-dimensional
spin space, with its main features: producing up to four quark generations,
two Higgs doublets, and both horizontal and vertical mass hierarchy
effects. The horizontal effect is implemented through non-Higgs fields/operators
for mass generation, which work analogously to flavons, and also
limit physics beyond the SM, as they are restricted by the (7+1)-dimensional space.

The mass relations obtained here  are not enough to accurately reproduce
the known quark masses on their own, but the model's features could
serve as a basis for further phenomenological studies, since they are derived and not postulated;
  e. g., in some models of dynamical symmetry breaking and compositeness\citep{BarShalom:2011zj,Hung:2010xh,Fukano:2011fp},
four-quark generations and two Higgs doublets are proposed at
the outset, whereas in the present case they arise naturally from
a minimum of physical assumptions. This is also true for the compositeness
aspect, which is naturally embedded in the present model by its very
construction.

As a SM extension, the extended spin-space model provides a unified description
of fields and symmetries, both gauge and spacetime, in a single matrix
space. It also provides an alternative unification scheme that does
not require a graded Lie algebra, as in SUSY. Thus, an interesting
research direction is to investigate to what extent the 
model serves as an alternative, or even a complement, to SUSY: this is a comparison
  fostered by the fact that both approaches can use Clifford algebras for their representations. 

 The model suggests directions to  research further,
such as a complete characterization of leptons, gauge-vector bosons,
and phenomenological implications for the CKM matrix and mixing
angles.  The features presented  thus far  make this a promising research
program, and because they are shared with   dynamical 
symmetry-breaking models, the model may provide new information
in this regard.
\begin{acknowledgments}
The authors acknowledge support from DGAPA-UNAM, project IN112916.
\end{acknowledgments}

\bibliographystyle{apsrev4-1}
\bibliography{tesis}

\end{document}